\documentclass[12pt]{iopart}
\usepackage{graphicx}
\usepackage{url}
\begin{document}
\title{Materials Design for New Superconductors}
\author{M. R. Norman}
\address{Materials Science Division, Argonne National Laboratory, 
Argonne, IL 60439, USA}
\ead{norman@anl.gov}
\begin{abstract}
Since the announcement in 2011 of the Materials Genome Initiative by the Obama administration,
much attention has been given to the subject of materials design to accelerate the discovery of
new materials that could have technological implications.  Although having its biggest impact
for more applied materials like batteries, there is increasing interest in
applying these ideas to predict new superconductors.  This is obviously a challenge, given that
superconductivity is a many body phenomenon, with whole classes of known superconductors
lacking a quantitative theory.  Given this caveat, various efforts to formulate materials design principles
for superconductors are reviewed here, with a focus on surveying the periodic table in an attempt
to identify cuprate analogues.
\end{abstract}
\submitto{\RPP}

\maketitle

\tableofcontents

\section{The Materials Genome Initiative}

In 2011, the Obama administration announced the Materials Genome Initiative \cite{mgi}.
As quoted that year in a presentation by Cyrus Wadia from the OSTP \cite{mgi2}, President Obama stated at 
the Carnegie Mellon University in June of 2011:
{\it To help businesses discover, develop, and deploy new materials twice as fast, we're
launching what we call the Materials Genome Initiative.  The invention of silicon circuits
and lithium ion batteries made computers and iPods and iPads possible, but it took years
to get those technologies from the drawing board to the market place.  We can do it faster.}
Of course, this did not come out of a vacuum.  For a number of years, practitioners of density
functional theory had been constructing databases, such as the Materials Project \cite{matproj},
which contains large numbers of electronic structure calculations which could then be data mined
for interesting properties.  This, combined with high throughput materials synthesis, could indeed
accelerate materials discovery \cite{depablo}.

Related to this has been the development of design principles for discovering new materials.
For instance, the properties of ferroelectrics are determined by atomic displacements
from high symmetry positions.  As such, these materials are well suited to density functional
methods.  As an example, Craig Fennie \cite{fennie} predicted that a high pressure form of
FeTiO$_3$ would be a multiferroic, and this material was subsequently grown by John Mitchell's
group and found to be so \cite{varga}.

Predicting a new ferroelectric is one thing, but a new superconductor is quite a different matter.
Even conventional superconductors involve a subtle interplay of electron-ion and electron-electron
interactions.  As such, predicting even the sign of the net interaction, much less the actual transition
temperature, is tough business.  Although it took only a few years from the advent of BCS theory \cite{bcs}
to the development of a quantitative strong coupling theory \cite{schrieffer}, this didn't help us much.
The theory did not predict the existence of even the simple material MgB$_2$ \cite{mgb2}, and even
after that discovery, attempts to predict new superconductors based on this class of materials, such as
Li$_x$BC, didn't pan out \cite{libc}.  Moreover, whole classes of unconventional superconductors, such as
rare earth and actinide heavy fermions, cuprates, and iron pnictide and chalcogenides, took the
community by complete surprise.  One can only imagine what's next.

Still, there have been several interesting attempts to apply both materials genome and materials design
principles to the discovery of new superconductors.  Here, these are reviewed, with some thoughts
about where the field is headed.

\section{Materials Genome Ideas Applied to Superconductors}

To the author's knowledge, there has been only one `hit' using materials genome ideas to predict a new
superconductor.  In 2010, Kolmogorov
and collaborators \cite{kolmo} used an evolutionary search procedure to identify new phases in the FeB
series.  For the several that were found, they then calculated the so-called Eliashberg function, which
is the phonon density of states weighted by electron-phonon matrix elements \cite{schrieffer}.  From this, they
predicted that a new orthorhombic FeB$_4$ phase would have a T$_c$ between 15 and 20K.  Several years later,
this material was synthesized and found to have a T$_c$ of 2.9K \cite{guo,ronning}.  Although perhaps not the
most spectacular success, it does demonstrate that this approach can work.  One issue is that even if
one has a good representation of the Eliashberg function, the actual value of T$_c$ is suppressed due
to Coulomb effects \cite{morel}.  The calculation of this repulsive $\mu^\star$ is on much less firm ground than the
attractive electron-ion interaction, though recent progress has been made \cite{bauer}.

Still, despite materials like MgB$_2$ with a T$_c$ of 40K \cite{mgb2},
T$_c$ of conventional superconductors tends to be limited
by retardation effects \cite{cohen}.  The true high temperature superconductors, at least at ambient
pressures \cite{h3s}, are cuprates and iron pnictides.  Here, we lack even a quantitative theory of what is
going on, though most feel that the attractive interaction leading to the formation of Cooper pairs is likely
due to magnetic correlations \cite{scalapino}.  Still, one could in principle develop descriptors of such materials,
and then use them to predict new superconductors.

Perhaps the best known attempt along these lines is that of Klintenberg and Eriksson \cite{klint}.
Based on a calculation of over 60,000 electronic structures, they then screened these to identify materials
with a predicted band structure similar to that of cuprates, that is, a quasi-2D material having a single d-p hybridized band
crossing the Fermi energy with a large hole-like Fermi surface around the $(\pi,\pi)$ point of the 2D Brillouin zone.
Several interesting candidates
were identified, such as Ca$_2$CuBr$_2$O$_2$, K$_2$CoF$_4$ and Sr$_2$MoO$_4$.
The author is unaware if any of these have panned out.  Moreover, there is a fundamental concern with such
an approach.  Doped layered manganites have a similar Fermi surface as cuprates, as seen
by angle resolved photoemission, yet the coherent ground state of this material is ferromagnetism, not
superconductivity \cite{sun}.  This means that the prediction is only as good as the descriptors, or expressed
more colorfully, one can easily fall into the GIGO (garbage in - garbage out) mode.

A related attempt has been reported by Curtarolo's group \cite{isayev}, again using structural motifs and
band structures as
descriptors, but this time looking at a variety of superconducting classes, including cuprates, pnictides, and
conventional superconductors.
Here, a more systematic approach was employed to identify the appropriate fingerprints within a given
class of materials, and then using this to predict T$_c$.  As can be imagined, there were materials whose T$_c$
was bang on, but others where the predicted T$_c$ was way off.  To date, if this has led to any new
superconductors, the author is unaware of it.  Still, this work and the work of Klintenberg and Eriksson is a start,
and perhaps with time, will lead somewhere.  Certainly, if in the future a `hit' emerges from these two papers,
more researchers will certainly pursue such endeavors.

\section{Superconductor Design Principles}

The formulating of design principles for superconductors has a long history.  Perhaps the most famous
advocate for this was Bernd Matthias.  His design principles were based on the then known class of highest
temperature superconductors, the cubic A15s like Nb$_3$Sn, and was colorfully summarized in a lecture by
Steve Girvin \cite{girvin}
\begin{itemize}
\item High symmetry is best
\item Peaks in the density of states are good
\item Stay away from oxygen
\item Stay away from magnetism
\item Stay away from insulators
\item Stay away from theorists
\end{itemize}
This is not completely fair.  Bernd was one of the early pioneers of non-conventional superconductors,
and in fact advocated looking for uranium-based superconductors proximate to a magnetic phase \cite{bernd1},
a prescient hint that eventually led to the discovery by others of unconventional superconductivity in UPt$_3$
and UBe$_{13}$ \cite{stewart}.  Still, the above design principles, based as they were on the cubic A15 compounds,
obviously represented the wrong direction when thinking about materials such as the cuprates,
which are quasi-2D doped magnetic oxide insulators.  And for sure, Bernd was highly suspicious of theorists,
quipping that the development of BCS theory did not lead to any increase in the discovery of new 
superconductors \cite{bernd2}.  On the other hand, the approach is certainly valid in that if the desired design principles are
correctly identified, then progress might be made.

\begin{figure}
\centering
\includegraphics[width=0.4\hsize]{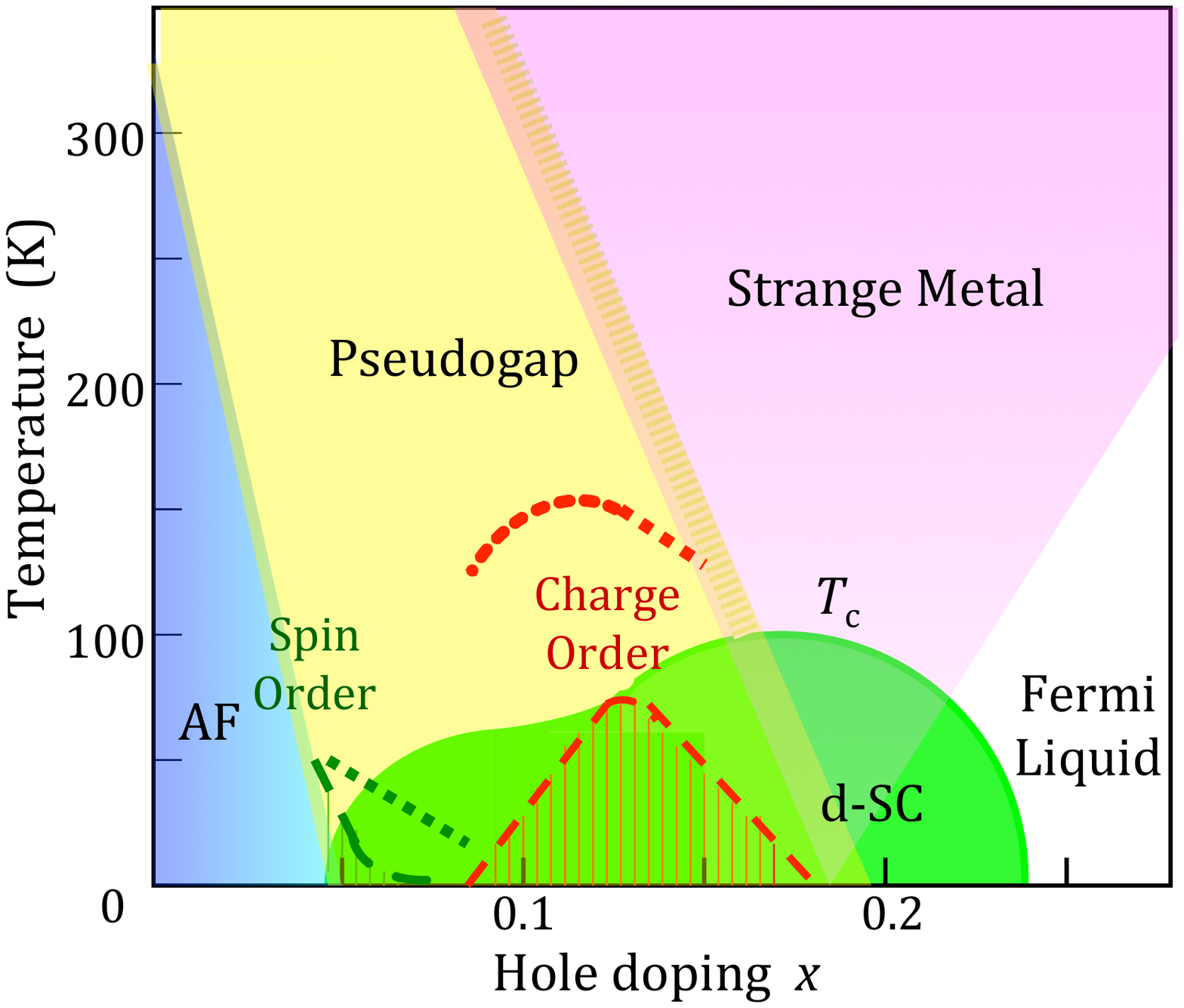}
\includegraphics[width=0.4\hsize]{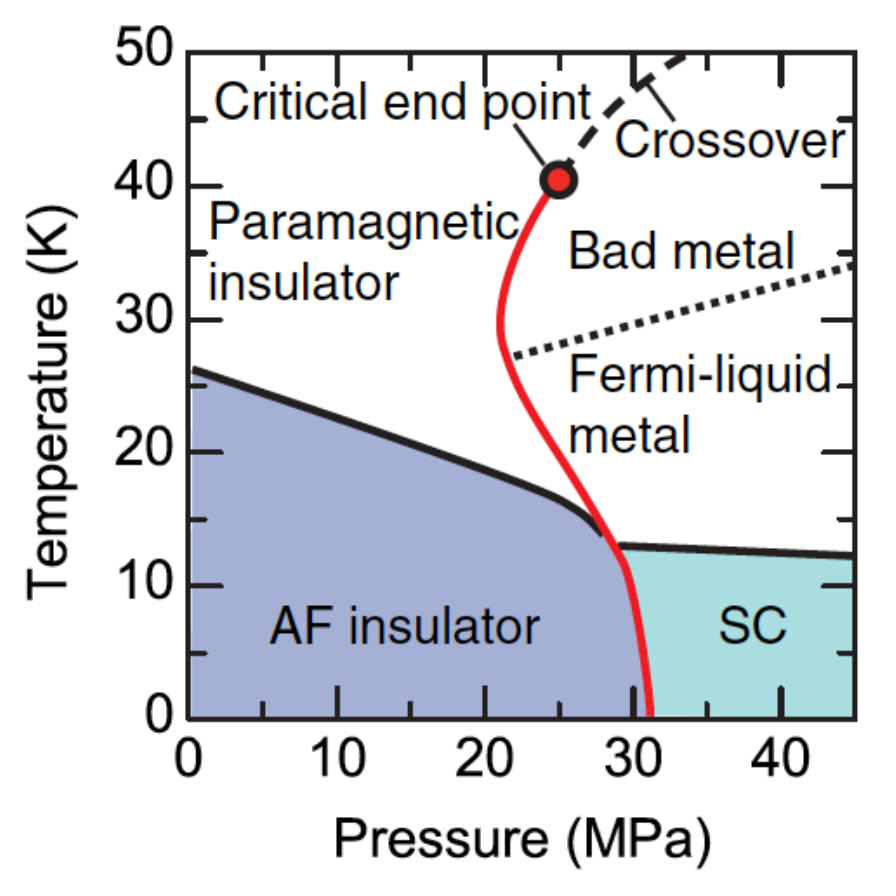}
\includegraphics[width=0.4\hsize]{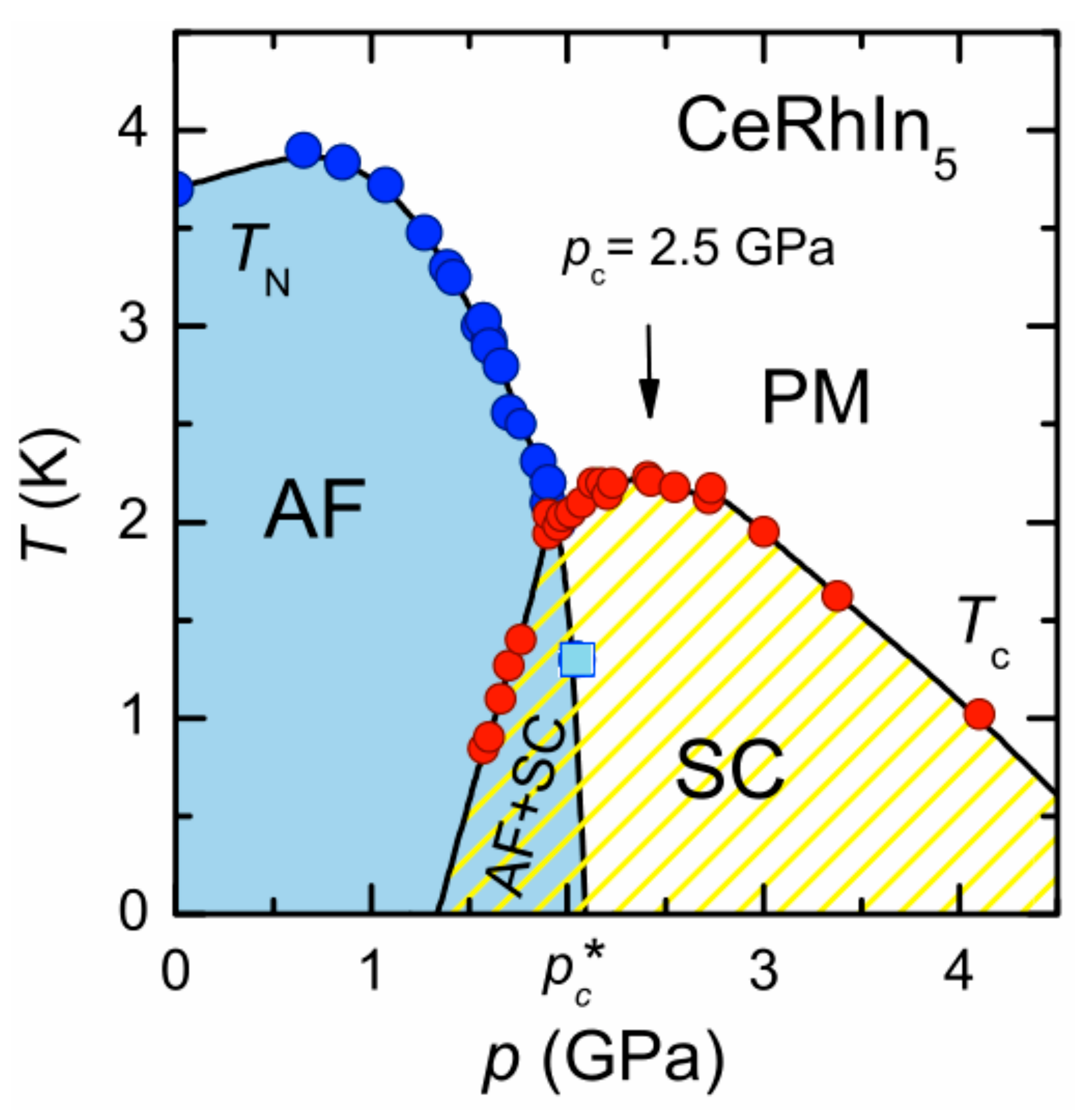}
\includegraphics[width=0.4\hsize]{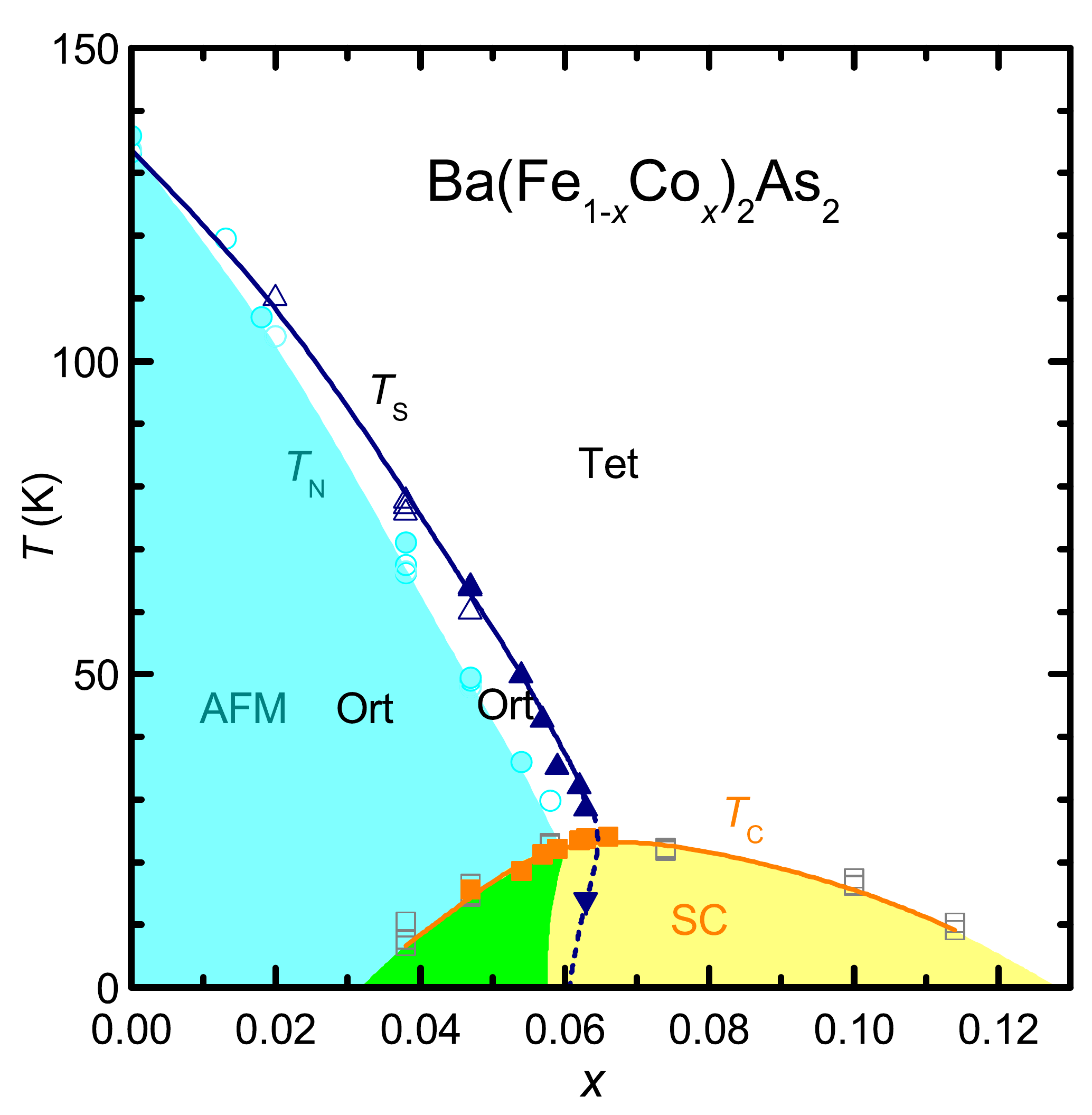}
\caption{Phase diagrams (temperature versus chemical doping or pressure) for four classes of superconductors: 
hole-doped cuprates like YBa$_2$Cu$_3$O$_{6+x}$
(upper left) \cite{keimer}, $\kappa$-(ET)$_2$Cu[N(CN)$_2$]Cl, a 2D-organic (upper right) \cite{furukawa}, heavy 
fermion CeRhIn$_5$ (lower left) \cite{knebel}, and an iron pnictide, Co-doped BaFe$_2$As$_2$ (lower right) \cite{nandi}.}
\label{fig1}
\end{figure}

Most unconventional superconductors are indeed found in proximity to a magnetic phase.  In Fig.~1, the phase
diagrams are shown for several classes
of superconductors: cuprates, pnictides, 2D-organics, and heavy fermions \cite{oxford}.
In all cases, the phase
diagrams are similar, including the not shown example of Cs-doped C$_{60}$ (buckyballs) \cite{takabayashi}.
One starts with a magnetic phase, typically an antiferromagnet (sometimes insulating,
sometimes not), and then uses a control parameter (such as chemical doping or pressure) to suppress the magnetic phase,
leading to a superconducting `dome' that eventually gets suppressed itself as the tuning parameter increases even further.
This design principle was realized early on by Gil Lonzarich's group and led to the discovery by them of heavy
fermion superconductivity in CePd$_2$Si$_2$ and CeIn$_3$ \cite{mathur}.  Most recently, it has led to the
discovery of superconductivity at high pressures in CrAs \cite{wu,kotegawa} and MnP \cite{cheng,norm15},
a very unusual occurrence given the strong magnetism exhibited by Cr and Mn.  Although T$_c$ of these
materials is small (2K and 1K, respectively), a quasi-1D variant, X$_2$Cr$_3$As$_3$ (X = K, Rb, Cs) has
been discovered with a higher T$_c$ of 6K \cite{cr3as3}.

This brings us to what I call the Goldilocks principle for high T$_c$.  Quasi-1D superconductors tend to have
low T$_c$ since fluctuations kill superconductivity in lower dimensions.  On the other hand, 3D materials
are limited in T$_c$ because interactions typically are weaker in higher dimensions.  So, 2D is just right, and sure
enough, the highest T$_c$ materials, cuprates and pnictides, are layered materials.

We will turn in the rest of this review to surveying the periodic table to see what might be worth exploring,
citing relevant work along the way \cite{oxford}.

\section{Looking for Cuprate Analogues}

\subsection{Cuprates}

\begin{figure}
\centering
\includegraphics[width=0.4\hsize]{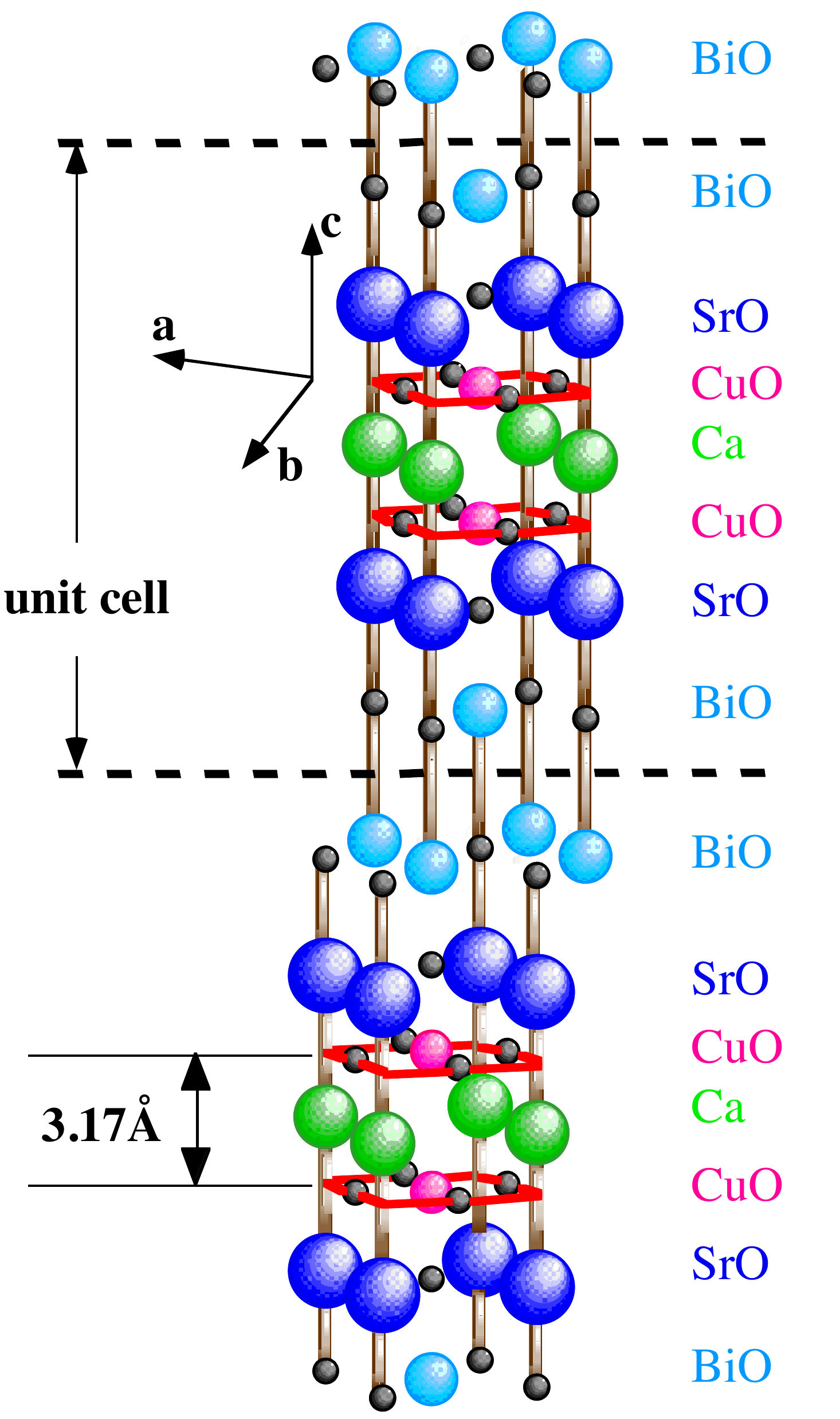}
\caption{Crystal structure of the cuprate Bi2212 (Bi$_2$Sr$_2$CaCu$_2$O$_8$).  Figure courtesy of Adam Kaminski.}
\label{fig2}
\end{figure}

Cuprates \cite{keimer} consist of CuO$_2$ layers (Cu being the metal, O the ligand), separated by spacer layers
that act to both isolate the active layers, and provide carriers to them to promote superconductivity (the stoichiometric
material being a magnetic insulator).  This spacer layer is typically composed of rare earths or other metals
such as Ca/Sr/Ba, or Hg/Tl/Pb/Bi, either in an isolated form
(where they separate CuO$_2$ layers in bilayer and trilayer variants), or liganded to oxygen (where
they separate successive CuO$_2$ layers or blocks of layers).  A good example is Bi2212, shown in Fig.~2.
The first thing to realize is the unique Jahn-Teller nature of the Cu$^{2+}$ ion, a design principle that led
Bednorz and Muller to the discovery of cuprate superconductivity to begin with \cite{rmp}.  This means that the
d$^9$ configuration of Cu has a single hole in the d x$^2$-y$^2$ Kramers doublet.
This suggests two possible pictures.  In the first, one forms a charge transfer insulator, where doped
holes reside in the oxygen orbitals, whereas doped electrons go onto the copper sites \cite{ZSA}.  In the doped
hole case, the holes on the four oxygen sites surrounding a copper site form a d x$^2$-y$^2$  configuration 
that reflects that of the 3d hole on the copper ion \cite{zhang}.  In a more
traditional band structure approach \cite{pickett}, the nearness in energy of the copper d x$^2$-y$^2$ and 
oxygen 2p states gives rise to a large bonding-antibonding splitting where the d x$^2$-y$^2$ state mixes with
the oxygen 2p$_x$ and 2p$_y$ states.  The resulting half-filled antibonding band then opens a correlation gap.
In both pictures, the Coulomb repulsion plays a fundamental role in creating the insulating state \cite{pwa}.
Carrier doping (for instance, by replacing say a 3+ ion in the spacer layer like La with a 2+ ion like Sr, thus donating holes
to the CuO$_2$ layers) then leads to a superconducting state as shown in Fig.~1.
In essence, cuprates can be thought of as doped Mott insulators \cite{leeRMP}, with the high T$_c$ thought
to be due to the extremely large superexchange interaction of 120 meV between the Cu ions mediated by the intervening 
oxygen ions \cite{pwa,leeRMP}.

Now, how does materials design enter?  One of the first attempts along this line was by Ole Andersen's
group \cite{pavarini}.  They noticed that T$_c$ scaled with the distance, d(Cu-apical O), between the copper and apical
oxygen atoms (the copper-planar oxygen separation does not vary much).  They realized that the p$_z$ orbital on the apical
oxygen, along with the 4s orbital on the copper site, helped to mediate longer range hopping in the
CuO$_2$ planes, in particular an effective hopping integral that acts between planar oxygen ions, denoted as
$t^\prime$, with $t^\prime$ and d(Cu-apical O) scaling together.  This line of
approach has subsequently been taken on by more sophisticated many-body techniques such as dynamical
mean field theory (DMFT).  Such studies \cite{weber} have shown a calculated correlation of T$_c$ with not only
these two parameters, but also with the separation of the energies of the d x$^2$-y$^2$ and the planar
oxygen 2p states (the smaller this energy difference, the higher T$_c$ is).  Recently, the same group has used an
evolutionary search algorithm to predict new copper oxysulfide variants that could be superconducting \cite{yee},
an idea that will surely be tested in the near future.

Another approach has been linked to the well known trend for bilayer cuprates to generally have
a higher T$_c$ than monolayer cuprates, and trilayer ones to have a higher T$_c$ than bilayer ones (though
layer numbers beyond three typically lead to a decreased T$_c$).  The accepted idea is that the electronic structure
of the inner and outer CuO$_2$ layers differ \cite{mukuda}, with one set of layers typically being underdoped
(and so having stronger correlations) and the other being overdoped (so, more metallic, and thus with a
presumably higher phase stiffness that suppresses fluctuations).  In general, such composite systems are
an ideal way of engineering T$_c$ \cite{berg}, and sure enough, by tuning the electronic structure of the inner
layer by pressure in trilayer Bi2223,
one can indeed enhance the T$_c$ of this material to 135K \cite{chen}.

From the beginning, though, the hope was that by adjusting the metal and/or the ligand ion, one might
achieve a new class of cuprate analogues (replacing oxygen by sulfur was already mentioned above \cite{yee}).
We begin with copper.  The superexchange $J$ is largest
in materials like those shown in Fig.~2 since the Cu-O-Cu bond angle is 180 degrees, which maximizes
this interaction (hybrid density functional based calculations have been able to give a good account of $J$
for a wide range of bond angles involving copper and oxygen \cite{rocque}).
So, you might ask, why would you want to decrease this angle?  The reason is that even
if one finds a lower T$_c$ analogue, this might represent a new class of materials.
A potential example is the quantum spin liquid Herbertsmithite, ZnCu$_3$(OH)$_6$Cl$_2$ \cite{shores}.  This copper hydroxychloride
mineral is composed of buckled Cu(OH)$_2$ layers, with
a Cu-O-Cu bond angle near 120 degrees, separated by Zn ions (with Cl ions playing the role of apical oxygens).
The important point is that the copper ions sit on a kagome lattice,
the most frustrated geometry for magnetism known in two dimensions.  In fact, the related triangular lattice was
the origin of Anderson's resonating valence bond (RVB) theory \cite{rvb} which subsequently led to one of the first attempts
to formulate a theory for the cuprates \cite{pwa}.  Along this line, Herbertsmithite appears to be a Mott insulator,
but does not exhibit magnetism down to the lowest measured temperature \cite{mendels}.  So, what if you could
dope it?  The success of this is not known (attempts so far have failed \cite{nytko}), but if you could, then
there is a prediction that by replacing Zn$^{2+}$ by Ga$^{3+}$ (which have almost the same ionic radius), one might 
obtain novel f-wave superconductivity due to the triangular nature of the lattice \cite{mazin}.  If so, then this
would be a cuprate analogue to the heavy fermion superconductor UPt$_3$, thought to be an f-wave
superconductor \cite{sauls} where the uranium ions also sit on a triangular lattice.

\subsection{Other 3d elements}

We turn now to replacing copper by another transition element.  Moving to the left along the periodic table, we
first come to nickel.  Interestingly, Sr-doped La$_2$NiO$_4$ exhibits the same charge stripes as exhibited by
underdoped cuprates \cite{tranquada,braden}.  But, the material remains an insulator.  Unlike the cuprates, where just
a few percent of dopants are needed to obtain mobile holes, the doping level in lanthanum nickelate has to exceed
100\% doping before a metallic state is achieved.  This metallic state looks remarkably like underdoped cuprates,
exhibiting a Fermi arc of gapless excitations centered around the Brillouin zone diagonal, but there is no evidence for
superconductivity \cite{uchida}.
This might be due to the valence of the nickel (with strontium doping, one is moving from
d$^8$ nickel to d$^7$ nickel (the cuprates being d$^9$), or the fact that one is far beyond the doping where
superconductivity would have occurred if the doped holes were not localized instead (in cuprates, superconductivity
typically only occurs between 5\% to 25\% doping).

There is, though, one potentially promising direction.  A trilayer variant, La$_4$Ni$_3$O$_8$, is known that has a 
crystal structure similar to that of electron-doped cuprates \cite{povalets}.  It is self-doped to 1/3 (hole) doping
relative to a d$^9$ configuration, making this a potential cuprate analogue.  This material, though, is an insulator,
and was recently discovered to have a similar charge stripe pattern to that exhibited
by 1/3 Sr-doped La$_2$NiO$_4$, that is La$_{2-x}$Sr$_x$NiO$_4$ with x=1/3 \cite{JZ1}.
Whether the doping in this 438 material could be altered from x=1/3 has yet to be demonstrated.

Similar physics occurs for the next element over, cobalt, with striped phases existing for Sr-doped La$_2$CoO$_4$
over a wide range of doping, analogous to that of the nickelates \cite{cwik,braden}.  Like the
underdoped cuprates, an unusual hourglass pattern is seen in the magnetic excitation spectrum ($\omega$
versus $q$), with an intermediate energy magnetic  `resonance' at $q$=($\pi,\pi$) at the neck of the hourglass \cite{boothroyd}.
But again, there has been no evidence for superconductivity, with the doped holes staying localized over a
wide range of doping.  On the other hand, there is a layered cobaltate, M$_x$CoO$_2$ (with M=Na, K, or Rb,
and intercalated by H$_2$O), where the cobalt ions sit on a triangular lattice, that does exhibit
superconductivity at around 5K \cite{takada}.  The nature of the superconductivity in this material is still
an active subject of debate, and whether a variant of this material could exhibit a higher T$_c$ remains to be seen.

\begin{figure}
\centering
\includegraphics[width=0.7\hsize]{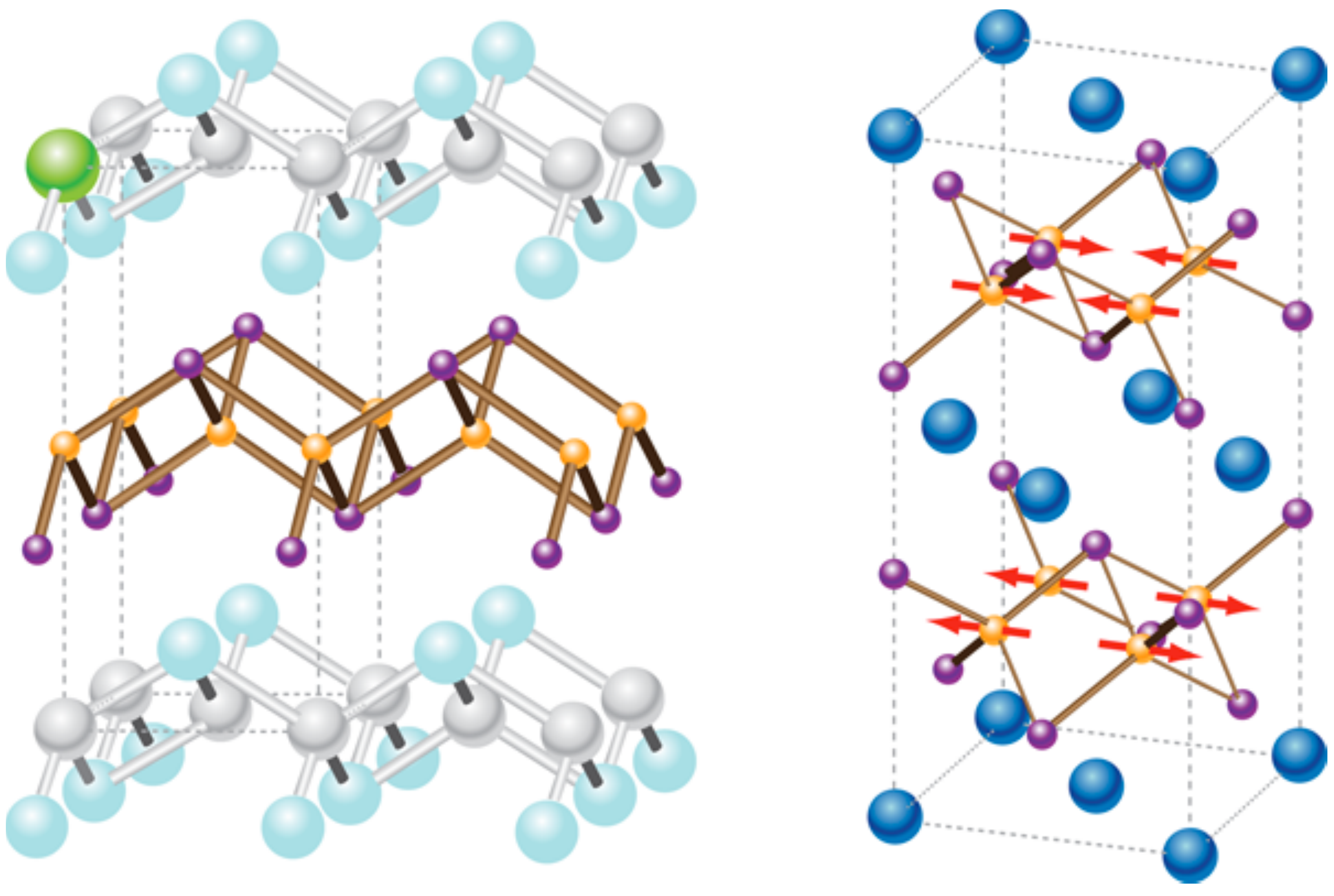}
\caption{Crystal structures of LaOFeAs (left) \cite{takahashi} and CaFe$_2$As$_2$ (right) \cite{goldman}
denoted as 1111 and 122, respectively. Yellow are iron atoms, purple are arsenic ones. On the left, a
fluorine dopant is shown in green. On the right, the spin directions (red arrows) on the Fe sites are shown for the
magnetic phase.}
\label{fig3}
\end{figure}

This brings us to the next element over, iron.  Had Hosono's group stopped after discovering 5K 
superconductivity in a layered iron phosphide \cite{FeP}, history might have been different.  But by playing
around with the ligand by replacing P by As, T$_c$ dramatically rose \cite{hosono}, leading to a new `iron age'
for high temperature superconductivity \cite{norm08}.  A variety of crystal structures exist, with the original 1111
material exhibiting superconductivity up to 56K \cite{56K}.  Other variants are known as 122, 111 and 11, with
the 11 variant actually involves S or Te as a ligand (FeSe, FeTe). 
Two examples (1111 and 122) are shown in Fig.~3.  As with the cuprates, these are layered structures,
but instead of being planar coordinated like the cuprates, the irons are tetrahedrally coordinated.
T$_c$ is very sensitive to the iron-ligand bond angle, exhibiting a sharp maximum when the ligand height
above the iron layer is near 1.38 \AA~\cite{mizuguchi}.  A proper description of these materials is well
beyond the scope of this article \cite{greg}, but again, the phase diagram of these materials is similar to
the cuprates (Fig.~1), and the superconductivity is thought to be driven by magnetic interactions \cite{chub}.  Unlike the
cuprates, multiple d-orbitals are involved near the Fermi energy, leading to a multi-band Fermi surface.
As a consequence, a sign changing superconducting order parameter
(necessary in magnetic mechanisms where the overall interaction is repulsive \cite{scalapino}) can be 
achieved by reversing the sign of that on the hole-like Fermi surfaces near $\Gamma$
relative to that on the electron-like Fermi surfaces at the Brillouin zone boundary \cite{pjh}.  This so-called s$_\pm$ state
is fundamentally different from the nodal d x$^2$-y$^2$ state exhibited by the cuprates, though the true nature
of the order parameter is not as clearly known as in the cuprates (depending on the doping and ligand, d-wave
and even conventional s-wave states have been advocated as well).

The most popularly studied iron pnictide is the 122 material (XFe$_2$As$_2$, with X=Ca, Sr, Ba) which can be doped by 
substituting on the Fe or As sites.  Surprisingly, this ThCr$_2$Si$_2$ structure is somewhat ubiquitous
for superconductors - the original heavy fermion superconductor CeCu$_2$Si$_2$ \cite{steglich}
exhibits this crystal structure, for instance.  Why certain crystal structures seem to be good for
superconductivity is still a murky area.  A classic example are the UX$_3$ materials \cite{koelling}.
Most of these have the same cubic A15 structure as mentioned earlier in the context of conventional
superconductors, yet are typically vegetables or magnets.  But there are two odd ball hexagonal
variants, UPd$_3$ and UPt$_3$, the former with a dHCP and the latter with an HCP crystal structure.
The first has an unusual quadrupolar order \cite{upd3}, the latter is a heavy fermion
superconductor \cite{stewart}.  As more such classes of materials are discovered, additional design
principles should become apparent \cite{isayev}.

\subsection{4d and 5d elements}

\begin{figure}
\centering
\includegraphics[width=0.7\hsize]{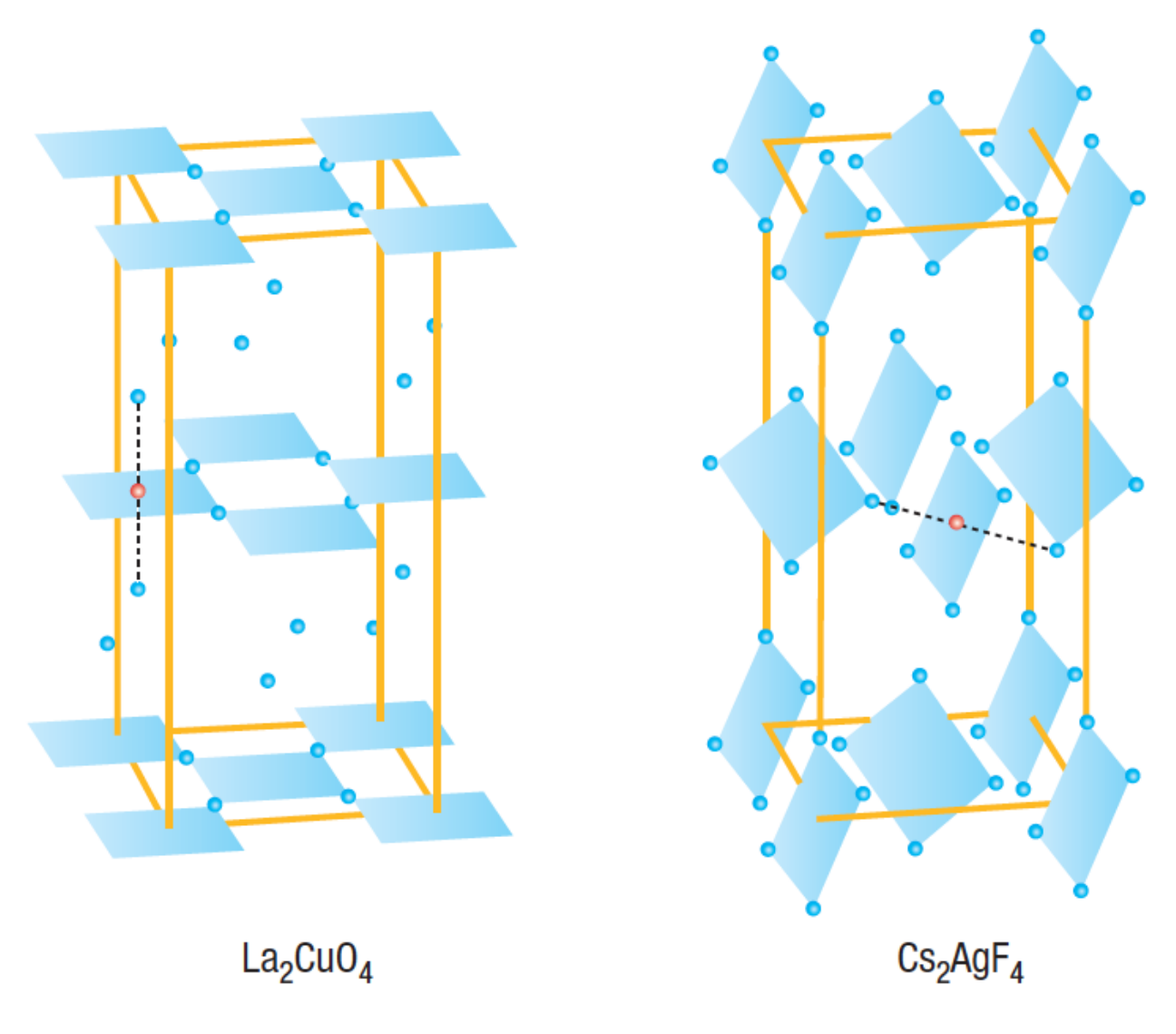}
\caption{Comparison of crystal structures of La$_2$CuO$_4$ (left) and Cs$_2$AgF$_4$ (right) \cite{grochala4}.
The shaded squares represent CuO$_4$ and AgF$_4$ plaquettes, with the apical axis shown as a dashed line.}
\label{fig4}
\end{figure}

Back to the cuprates, what should happen if we move down to the next periodic row instead?  The
element below copper is silver.  One issue with silver is that it typically is a valence skipper, primarily
forming either Ag$^+$ or Ag$^{3+}$.  But under certain conditions it can be stabilized as Ag$^{2+}$.
In that sense, fluorine is a more useful ligand than oxygen \cite{grochala1}.  Energetically, the fluorine
2p energies are intermediate between Ag$^{2+}$ and Ag$^{3+}$ as oxygen is between Cu$^{2+}$
and Cu$^{3+}$, leading to similar charge transfer physics \cite{grochala2}.  The big problem is getting
these fluoroargentates to form the desired layered structure \cite{grochala2}.
One issue is that Ag$^{2+}$ often exhibits an inverse Jahn-Teller configuration (with the d$^9$
hole in the 3z$^2$-r$^2$ orbital instead), and even if found in a Jahn-Teller configuration, it often
responds by forming AgF$_4$ units whose normals are not along the $c$-axis (Fig.~4) \cite{mazej}.  Finding
the optimal `cuprate' configuration (flat AgF$_2$ sheets with a long apical axis parallel to $c$) is a difficult
challenge that has yet to be realized \cite{grochala2}.  In that context, a recent set of density functional calculations
were done to predict superlattices that would stabilize such a structure \cite{agf2}.  Although synthesizing
these materials could prove to be difficult, a success here could start a whole new field
of superconductivity \cite{grochala3}.
Now, there are oxides of silver which are low T$_c$ superconductors, such as
Ag$_7$O$_8$HF$_2$ \cite{kawashima}, but these clathrates' superconductivity likely originates
from `rattling' modes of HF$_2$ centered in the large Ag$_6$O$_8$ cages of this material.

The next element over in the 4d row is palladium.  As with nickelates, Pd$^{2+}$ is d$^8$ rather than d$^9$
and so is expected to differ from cuprates.  On the other hand, Pd metal is nearly ferromagnetic, and so
Pd compounds can be expected to have interesting magnetic properties.  And if the magnetism is
suppressed, who knows, perhaps some kind of superconductivity might be observed.  The prediction of
p-wave superconductivity in Pd goes back a long time \cite{fay}, and though not realized (yet), was shortly
thereafter co-opted to be one of the leading theories for triplet superfluidity in $^3$He.  In that context,
the recent observation of ferromagnetism in Ba$_2$PdO$_2$Cl$_2$ is of note, this material having a
similar crystal structure to that of electron-doped cuprates \cite{tsujimoto}, so doping studies could be
of interest. In connection with the iron
pnictides, superconductivity at 3.9K was recently discovered in BaPd$_2$As$_2$, possessing the
same ThCr$_2$Si$_2$ crystal structure as BaFe$_2$As$_2$ \cite{guo-ren}.

No further remarks will be made about the 4d row, expect to note that low temperature, probably p-wave,
superconductivity has been found in Sr$_2$RuO$_4$ \cite{maeno}, with its bilayer variant, Sr$_3$Ru$_2$O$_7$
\cite{mckenzie}, exhibiting electronic nematic order as also seen in the cuprates and iron pnictides.

We now jump down to the very interesting 5d row.  Although a variety of Pt and Au compounds have similar
formula units to their cuprate counterparts, the discussion here is confined to two 5d elements,
Os and Ir.  Os is interesting in that it exhibits a variety of valence states, and so there is a vast materials space
here remaining to be explored.  A magnetically frustrated pyrochlore,
KOs$_2$O$_6$, exhibits low temperature superconductivity, but this is likely due to K `rattling' modes, making
the material similar to the AgO clathrate that was mentioned earlier \cite{yonezawa}.

\begin{figure}
\centering
\includegraphics[width=0.4\hsize]{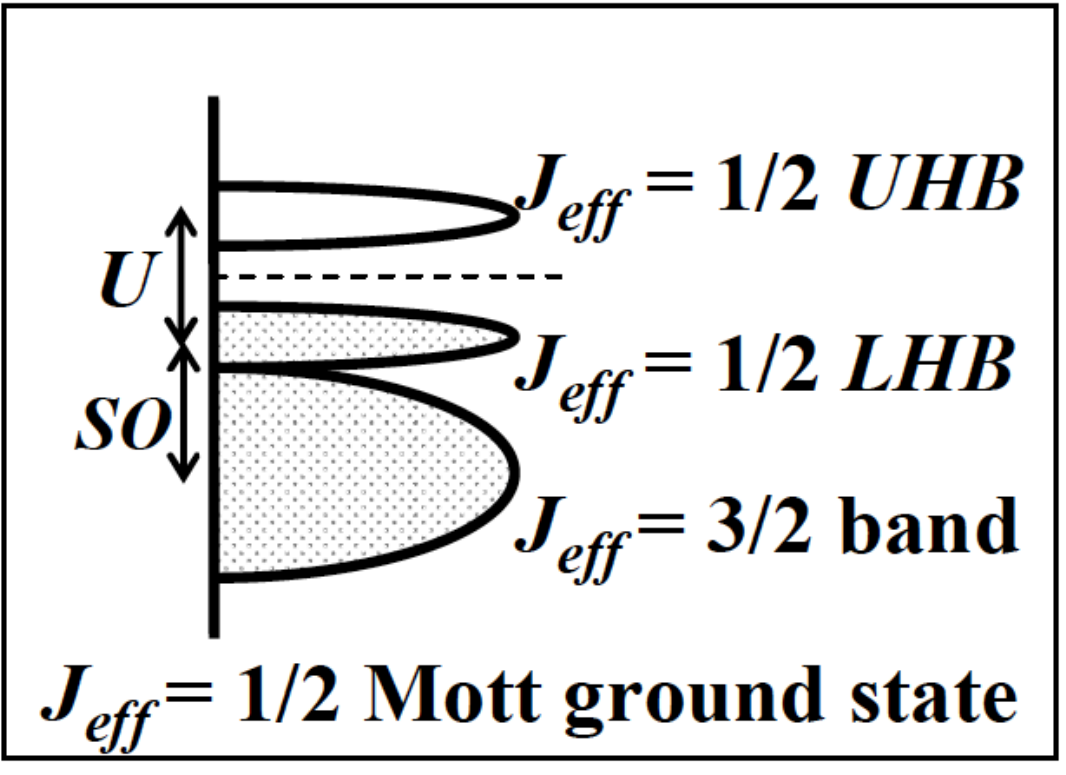}
\caption{Electronic structure of Sr$_2$IrO$_4$ \cite{kim1}.  $U$ is the on-site Coulomb repulsion, $SO$ the spin-orbit
splitting, with $LHB$ and $UHB$ denoting the lower and upper Hubbard bands.  The dashed line is the
chemical potential.}
\label{fig5}
\end{figure}

Of perhaps greater interest is iridium.  Sr$_2$IrO$_4$ exhibits a similar electronic structure to cuprates despite
its d$^5$ configuration (Fig.~5).  This is due to the profound effect of spin-orbit, which splits the t$_{2g}$ manifold into
a filled quartet and a half-filled doublet \cite{kim1}.  Although there are definite differences (the
iridates are likely not charge transfer insulators due to the smallness of the Hubbard $U$ and the energetic
separation of the Ir 5d and O 2p states), still, the presence of a half-filled Kramers doublet that opens up a correlation
gap is reminiscent of the cuprates (and the stoichiometric material is an antiferromagnetic insulator as well,
just like the cuprates).  This has led to the prediction of unconventional superconductivity \cite{wang}.
Unfortunately, iridates, having Sr in its spacer layer rather than La, have proven to be difficult to chemically dope.  On the other
hand, surface sensitive studies have made some progress along these lines.  By depositing potassium on the surface
to dope it (a `trick' that was earlier used for cuprates \cite{andrea}), B J Kim and collaborators were able to observe Fermi
arcs in iridates, very similar to what was seen in underdoped cuprates \cite{kim2}.  More recently, by introducing bilayer
inclusions (which are metallic), they have been able to exploit their screening to go to low temperatures and
observe a d-wave energy gap \cite{kim3} that has also been seen by STM \cite{stm}.  This leads to the hope
that an appropriately doped sample might exhibit superconductivity.

This brings us to an important point.  Although much theoretical work has been devoted to ideal materials,
there has been much less attention given to doping.  This is obviously an issue, since most of the interesting
materials like cuprates, pnictides, and buckyballs, are rarely superconducting in their stoichiometric phases,
and require some form of chemical doping to achieve superconductivity.  Many interesting stoichiometric
materials, like the above example of iridates, have proven difficult to dope.  And others, once doped,
often exhibit charge localization (nickelates and cobaltates being relevant examples).  Earlier, the
interesting case of Herbertsmithite was mentioned, which was predicted in its doped state to be a
novel superconductor \cite{mazin}.  Recently, a density functional study was done to investigate doping
3+ and 1+ ions on the Zn$^{2+}$ interlayer sites, with the claim that a number of such ions are promising
candidates \cite{guterding}.  On the other hand, synthetic techniques have so far failed \cite{nytko},
likely due to the fact that in reality, these copper hydroxychloride minerals can only accommodate 2+ ions on
the interlayer sites (otherwise, they fall apart).  Therefore, much more challenging work remains to be done, both
experimentally and theoretically, before we get a handle on the crucial issue of chemical doping.

\section{Exploiting Layering}

The above mentioned superconductors are not the only layered ones.  There is a class of layered nitrates
that have T$_c$ up to 26K, HfNCl and its siblings (ZrNCl, TiNCl) \cite{yamanaka,kasahara}.
These are ionic band insulators,
and although the electron-ion interaction is likely prevalent, other interactions also come into play \cite{yin}.
Doping is achieved by intercalation (such as with Li), but so far, only electron doped materials have been studied (with 
the doped electrons entering the empty d bands).  Hole doping, where the doped holes enter into the
ligand 2p states (as in the charge transfer picture for cuprates), would be interesting.
In that context, the stoichiometric material has also been made superconducting by liquid ion gating \cite{ye-iwasa},
and so one could easily reverse the gating potential and see what happens.

This brings us to a vast new field where in this short article, justice cannot be given.  This is the study of engineered
2D materials.  This takes two forms.  First, superlattices of materials can be constructed by MBE growth.
There have been a number of successes here.  In cuprates, one can take blocks of material which are
respectively undoped insulators and heavily overdoped non-superconducting metals, and then at their interface,
high temperature superconductivity can be achieved \cite{bozovic}.  Or, one can take a single layer on a substrate and gate
it as mentioned in the previous paragraph, leading to superconductivity in a variety of 2D materials \cite{iwasa}.
Perhaps the most spectacular example is the recent claim
of superconductivity near 100K in a single layer of FeSe on top of a substrate \cite{ge}.

Related to this has been the development of a new modular design principle that treats layers (or multilayers) much
like Lego blocks that can be interchanged to engineer new materials.  Several years ago, this
was used to predict the existence of a complex iron oxide phase,
Y$_{2.24}$Ba$_{2.28}$Ca$_{3.48}$Fe$_{7.44}$Cu$_{0.56}$O$_{21}$, related to that of Bi2212 \cite{dyer}.
The synthesized structure was close to the predicted one.
More generally, most oxides are octahedrally coordinated, and given their importance for a variety of
desired physical properties (ferroelectricity, magnetism, etc.), design rules have
been developed known as `octahedral engineering'  \cite{rondinelli}.  The impact of both of these works on
superconductivity could be realized in the near future.

\section{Some Final Thoughts}

Just because one has discovered a high temperature superconductor does not mean that technologically, it
would have any relevance.  A trivial example is the recent spectacular discovery of ultra-high temperature 
superconductivity (near 200K) in H$_3$S under extremely high pressures \cite{h3s}.  Such high pressures are 
needed to metalize the hydrogen (sulfur acting as a matrix to help bring the needed pressure down), with
the prediction that metallic hydrogen would be superconducting due to its high energy phonon modes being
made a long time ago \cite{ashcroft}.  But a more relevant example are the layered materials we have been
focusing on in this review.  In two dimensions, a true superconducting transition does not exist, rather one
gets a Kosterlitz-Thouless transition \cite{kosterlitz}.  3D coupling between the layers stabilizes
3D superconductivity, but as this coupling is weak, it has a weak effect on pinning vortices.  So, the combined
effect of fluctuations (always relevant below 3D) and weak pinning is detrimental for obtaining high
critical currents \cite{beasley}.  This is one of the reasons that cuprates have made a small impact so far on technology despite
their high critical temperatures.  Recent progress has been made, though, on designing pinning landscapes
to optimize superconductors, both for cuprates and for iron pnictides, leading to a `materials genome' for defect 
engineering in superconductors \cite{kwok}.  That is, designing superconductors requires not only tuning
the electronic structure and interactions (via the choice of metal, ligand, and crystal structure), but also by
tuning the carrier concentration (chemical doping or pressure) and the defect landscape (to enhance pinning).  It is
only by thinking about all three of these facets will we have a hope of designing ideal superconductors.

I end, though, with a cautionary note.  The history of superconductivity has been littered by bad ideas.
Perhaps the most colorful description of this has been given by Bernd Matthias in a number of talks, that
although over forty years ago now, still seem highly relevant today:
\begin{itemize}
\item {\it The electron-phonon interaction always reminds me of the man who is looking for his keys under a 
street light and his friends say ``but you didn't lose them here, you lost your keys over there''. ``I know, but it 
is too dark over there.''} \cite{bernd3}
\item {\it Sometimes this thing reminds me of the Virginia Wolf play where four people argue all night about the aberrations of a 
child, and when the play is over, it turns out that there never was a child.  This is exactly how I feel about the 
organic superconductors.} \cite{bernd3}
\item {\it That of course leads you to Green's functions and the absence of any further predictions.} \cite{bernd4}
\item {\it Unless we accept this fact and submit to a dose of reality, honest and not so honest speculations will 
persist until all that is left in this field will be these scientific opium addicts, dreaming and reading one another's 
absurdities in a blue haze.} \cite{bernd5}
\end{itemize}
Who knows, perhaps he would have had bad thoughts about a `material genome' approach to superconductivity
as well.  Certainly, there is a lot of skepticism in the field about reliable predictions concerning superconductivity based
on electronic structure simulations \cite{anderson}.  After all, almost all of the great discoveries in the field have been
made serendipitously \cite{mandrus}.  Still, until such predictions can be made, then we will not truly understand
the phenomenon of superconductivity in its many varied forms.  This is indeed a noble goal to strive towards.

\ack

This work was supported by the Center for Emergent
Superconductivity, an Energy Frontier Research Center funded by the
US DOE, Office of Science, under Award No.~DE-AC0298CH1088.
The author thanks many of his colleagues for discussions, that
have helped form the opinions expressed here.

\end{document}